\documentclass{ws-procs9x6}

\begin{document}

\title{The 6C** Sample and the Highest Redshift Radio Galaxies}

\author{M.~J. Cruz, M.~J. Jarvis, K.~M. Blundell and S. Rawlings}

\address{Oxford University Astrophysics, Denys Wilkinson Building,\\
  Keble Road, Oxford, OX1 3RH, United Kingdom\\ 
}


\maketitle

\begin{figure}[ht]
  \begin{center}
    \begin{minipage}[l]{0.49\linewidth}
      \epsfxsize=2.3in\epsfbox{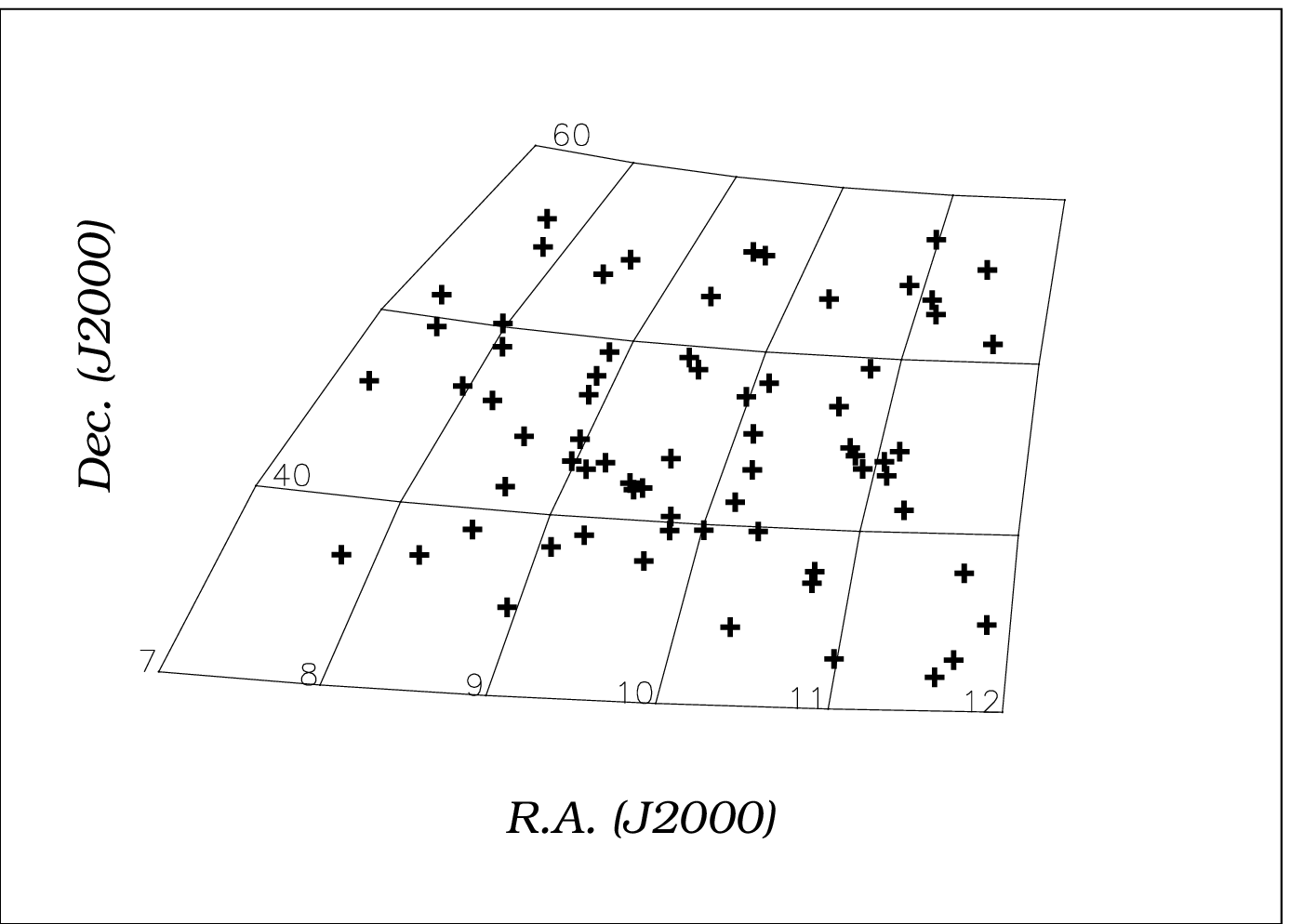}   
    \end{minipage}\hfill
    \begin{minipage}[tl]{0.49\linewidth}
      \epsfxsize=1.9in\epsfbox{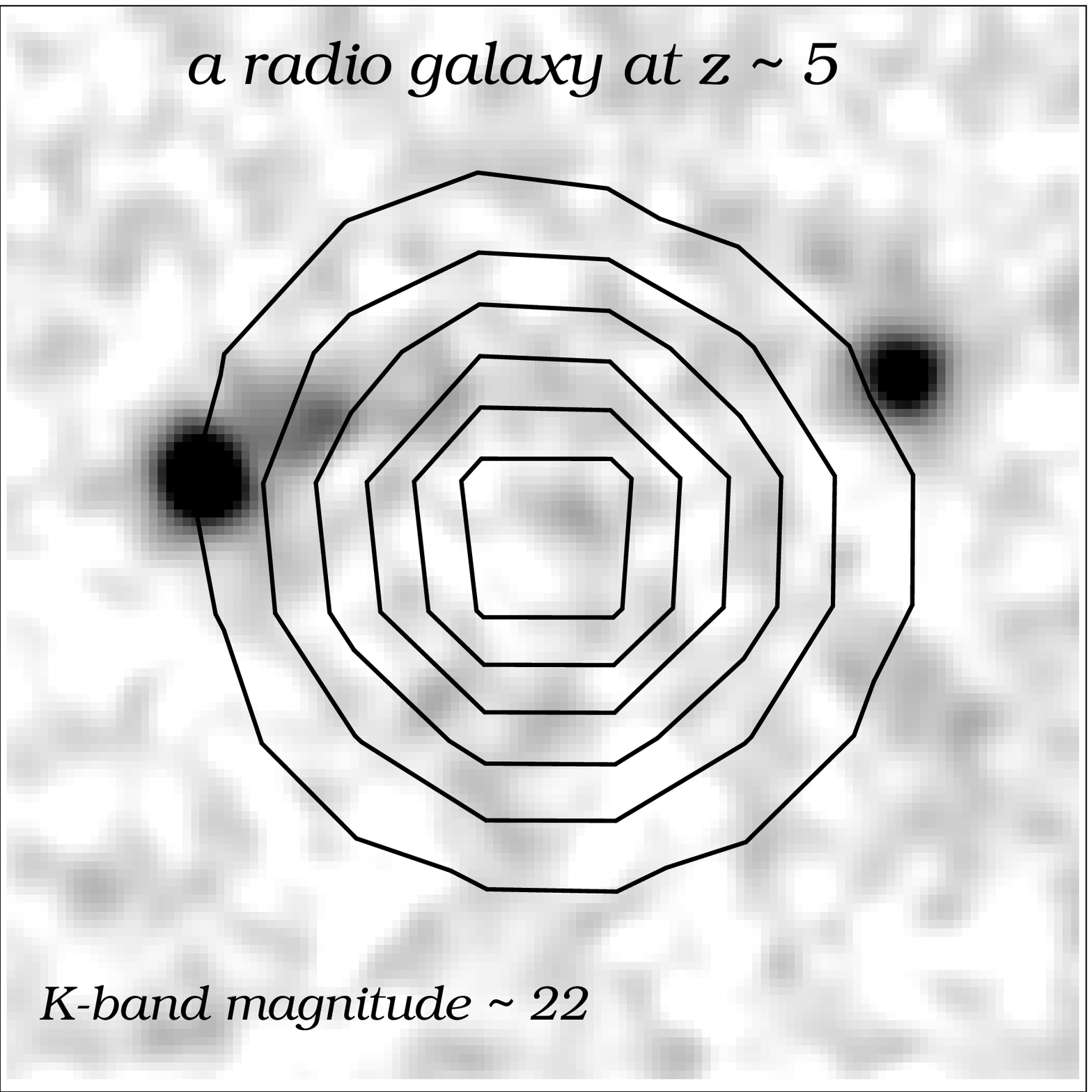}   
    \end{minipage}\hfill
  \end{center}
\caption{ {\bf Left} Location of all 6C** sources within the survey
  region. {\bf Right} The 1.4\,GHz radio contours overlaid onto a NIRI
  (Gemini) K-band image of a candidate $z > 5$, 6C** radio
  galaxy. The image has a size of $14\,^{\prime\prime}$ x $18\,^{\prime\prime}$.}
\label{fig}
\end{figure}

\abstracts{We present a new radio sample, 6C** designed to
  find radio galaxies at $z > 4$ and discuss some of its near-infrared 
 imaging follow-up results.}
\vskip -0.1in
\noindent
{\bf Why search for the highest redshift radio galaxies?} Radio
galaxies trace the most massive galaxies\cite{jarvis2001b}
($>2\rm{L}_{*}$) and are
associated with the most massive black holes\cite{willmclujar}
($\approx 10^{9}\,\rm{}M_{\odot}$) in the
universe at every epoch. 
Recent studies support the idea that at $2 <z <4$ they reside in
proto-clusters and are progenitors of the central brightest cluster 
galaxies\cite{kurk03}. The highest redshift radio galaxies ($z > 4$)
are therefore key targets for studies of formation and evolution of
massive structures in the early universe. They are particularly useful
in this respect as they are selected on the basis of their radio
emission and thus free of problems associated with optical selection
methods such as dust obscuration. Also, they tend not to have their 
optical and infrared emission dominated by non-stellar nuclear
emission as is the case for quasars.
\noindent
\newline
\newline
\noindent
{\bf The 6C** Filtered Sample}
\noindent
This is a new sample of radio galaxies drawn from the 151\,MHz, 6C
survey which has been filtered with radio criteria chosen to optimize 
the chances of finding radio galaxies at $z > 4$. 
It has been selected to be brighter than 0.5\,Jy at 151\,MHz on an area
of sky of 0.33\,sr and to exclude sources whose radio spectral index
between 151\,MHz and 1.4\,GHz are flatter than 1 or whose radio
angular size are larger than 12 arcsec. 
These are characteristics invariably seen in very distant 
radio galaxies\cite{mccarthy93,rawlings96,vB99}. The selection
criteria resulted in the 6C** sample comprising 69 objects; their
location within the survey region is shown in Fig.~\ref{fig} (left).
Based on the work of Ref.~\refcite{jarvis2001c} we expect to have
at least two sources at $z > 5$ among them. 
\noindent
\newline
\newline
{\bf Results and Discussion}
\noindent
Deep K-band imaging follow-up with UFTI on UKIRT and NIRI on Gemini
has provided us with near-infrared identifications for every member in our
 sample. K-band photometry provides an accurate method of redshift
 estimation by using the tightness of the $K-z$ diagram\cite{willott03}. 
We estimate that $\sim$ 40 \% of the sources on 6C** have redshifts
$>$ 2, in accordance with extrapolations from previous  
studies\cite{jarvis2001a}.
By selecting the faintest ($K \sim 21$) members on our sample we have 
identified five strong candidate $z > 5$ radio galaxies. One of them
was not securely detected despite a 45 min. integration with 
NIRI, although there are hints of an object with $ K \sim 22$ close to
the limit of the observation (Fig.~\ref{fig}, right). Future spectroscopic
observations will tell us about the nature of these sources and will
secure their redshifts.
\vskip -0.1in
\phantom{x}
\noindent
{\bf Acknowledgments}
\noindent
{\small MJC acknowledges the support from the Portuguese 
 Funda\c{c}\~{a}o para a Ci\^{e}ncia e a Tecnologia, 
and Corpus Christi College,
 Oxford.}
\vskip -0.3in
\phantom{x}

\end{document}